# Can we start sharing our rides again? The postpandemic ride-pooling market

Olha Shulika[a] and Rafal Kucharski[b]

*Faculty of Mathematics and Computer Science, Jagiellonian University, Krakow, Poland*

[a] olha.shulika@uj.edu.pl

[b] rafal.kucharski@uj.edu.pl

Before the pandemic ride-pooling was a promising emerging mode in urban mobility. It started reaching the critical mass with a growing number of service providers and the increasing number of travellers (needed to ensure ride-pooling efficiency and sustainability). However, the COVID pandemic was disruptive for ride-pooling. Many services were cancelled, several operators needed to change their business models and travellers started avoiding those services. In the postpandemic period, we need to understand what is the future of ride-pooling: whether the ride-pooling system can recover and remain a relevant part of future mobility.

Here we provide an overview of the postpandemic ride-pooling market based on the analysis of three components: a) literature review, b) empirical pooling availability survey and c) travellers' behaviour studies.

We conclude that the core elements of the ride-pooling business model were not affected by the pandemic. It remains a promising option for all the parties involved, with a great potential to become attractive for travellers, drivers, TNC platforms and policymakers. The travel behaviour changes due to the pandemic seem not to be long-lasting, our virus awareness is no anymore the key concern and our willingness to share and reduce fares seem to be high again. Yet, whether ride-pooling will get another chance to grow remains open. The number of launches of ride-pooling start-ups is unprecedented, yet the financial perspectives are unclear.

## 1 *Introduction*

Ride-pooling is a flexible on-demand car travel service that makes people share their trips simultaneously with other co-travellers (Figure 1). Two or more travellers request their travel demand to the platform, which matches their request into a single, feasible pooled route and assigns a vehicle to serve it. The vehicle serves as such a pooled ride picking up consecutive travellers at their origins and travels with them towards their consecutive destinations. Faced with detours, wait times and potential walks to stops, pooling travellers are incentivized to pooling with a reduced trip fare compared to a private ride. This reduction shall compensate for the discomforts of pooling to be attractive for travellers. Which requires either substantial discounts for pooled rides or a big demand level, to guarantee high compatible matches among pooled travellers.

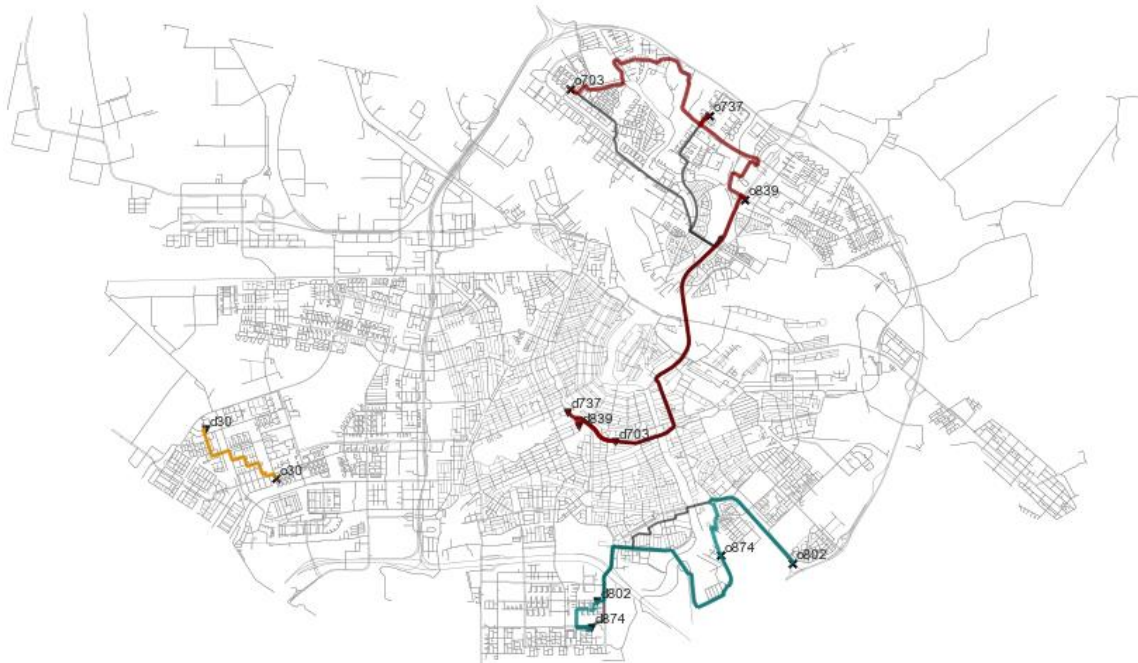

Figure 1. Example of pooled trips in Amsterdam: a) non-shared, private ride marked yellow in the West, b) ride shared by two travellers marked green in the South and c) ride shared by three travellers marked brown in the North. Stars denote origins, triangles destinations and grey bold lines travellers' shortest paths respectively (Kucharski & Cats, 2020).

Ride-pooling is a promising emerging mode in urban mobility. It has the potential to improve mobility, especially in those areas that are not covered by public transport (assisting riders with first-mile/last-mile issues) (Abouelela et al., 2022; Atkinson-Palombo et al., 2019; König & Grippenkoven, 2020; Mohamed et al., 2019). Ride-pooling may encourage car-free access to transit stations (Abdullaha et al., 2020, 2021), allows providers to diversify their services and expand their market towards new segments and opens opportunities for additional profits. In turn, travellers, naturally maximising the utility of their trips, may be attracted to the ride-pooling service due to its greater flexibility compared to public transport and lower cost with a slight increase in travel discomfort compared to ride-hailing (Atasoy et al., 2015; König & Grippenkoven, 2020).

This organisational innovation, despite being highly challenging to implement and attracting sufficient market shares, is beneficial for all the parties involved. Ride-pooling involves travellers - willing to share their rides; drivers - willing to work for the platform and on-demand mobility platforms (Transportation Network Companies (TNCs)) - matching co-travellers with available drivers (Dandl et al., 2021). The development of a ride-pooling service is possible since the interests of all participants are aligned: travellers want to reduce their trip fares, drivers want to serve more demand, TNCs want to better utilise their fleet and, notably, the policymakers view ride-pooling as potentially sustainable, when high occupancies allow reducing vehicle miles. Such alignment was the main driver behind the rapid rise of ride-pooling in a pre-pandemic world.

*1.1. Research questions*

Here we ask whether ride-pooling remains attractive in the postpandemic world and whether it can recover and gain the momentum it had in the pre-covid era. In this study we explore the changes in the interaction between ride-pooling market participants; analyse the availability of ride-pooling service and provide a general postpandemic ride-pooling market trend. To do this, we report the changes in the composition of participants in the ride-pooling market composition in the postpandemic period; consider the political aspects of the public sector that regulate the activities of ride-pooling operators; we analyse the declared policy of work of ride-pooling providers in the postpandemic period; we analyse the empirical pooling availability of services in more than 20 major cities around the world, where providers provided ride pooling services before the pandemic; we also analyse the general aspects of the postpandemic ride-pooling market, which are the main drivers for reaching the critical mass with a growing number of service providers and an increasing number of travellers.

In general, such study is highly challenging, yet equally important. It is timely since the ride-pooling is just about to restart after the pandemic. However, there are no scientific materials at hand. The studies either rely on pre-pandemic trends or report the pandemic disruptions and are hardly representative of the global ride-pooling market. Nonetheless, we decided to provide a comprehensive overview using available materials. Contrary to the classic reviews, we relied also on the press releases, marketing materials and interviews with the ride-pooling professionals. Such compilation of available sources, complemented with empirical research on the actual availability of ride-pooling provides a broad picture needed to conclude the future of ride-pooling.

To make this goal even more challenging, the ride-pooling concepts evolved during the pandemic, i.e. the para-transit, on-demand flexible public transport is often seen as a more promising and sustainable alternative to ride-pooling (which is reflected also in the number of new companies launched recently to offer on-demand transit (Foljanty, 2021). Here, we focus on ride-pooling offered by TNC companies and acknowledge that it is likely that in future it will be replaced by more compact, on-demand transit-like services.

*1.2. The brief history of the rise and fall of ride-pooling*

*1.2.1 Early concepts*

In the early stages, pooling services were more of informal and unorganised activity, historically denoted carpooling. Carpooling was a non-profit common trip service where drivers offer vacancies in their cars while passengers share the cost of the trip with the driver (Anthopoulos & Tzimos, 2021). It appeared in the USA during World War II as a more economical use of private vehicles for personal transportation and it attracted public attention during periods of fuel crises as a lever for saving fuel resources until the late seventies of the twentieth century (Ferguson, 1997). However, even carpool demonstration projects funded by the federal government failed to prove viable in the eighties and nineties. The increased number of people moving to the suburbs and the disaggregation of trip destinations scattered randomly across suburbs yielded more complex travel patterns, which eventually made carpooling less relevant (Rosenbloom & Burns, 1993).

This early carpooling idea was brought back to life in the form of ride-pooling shortly after the rise of the ride-hailing platforms. Due to computational challenges, it quickly became obvious that the real-size ride-pooling systems have to be formalised and assisted with mathematical optimization methods, like the shareability networks, which allowed to efficiently compute optimal sharing strategies on massive datasets (Santi et al., 2014). In parallel, digital information and communication technologies have allowed the widespread use of smartphones and real-time online applications, thereby facilitating the rise of platform-based ride-hailing taxi services and paving the way toward pooled mobility (Furuhata et al., 2013).

*1.2.2 The rapid growth*

Mass adoption of information and communication technologies gave good momentum to the rapid development of app-based on-demand ride services (TNCs), which began to provide a wide variety of real-time and demand-responsive trips (Baker, 2020). Quick and efficient matching of supply (vehicles and drivers) and demand (travellers) allowed TNCs to offer alternatives not only for long-distance pooled trips but also for pooled trips on short distances (Friedrich et al., 2018).

The expansion of the number of Internet-enabled mobile devices enabled dynamic ridesharing - a service that automatically matches drivers and riders close to their desired departure times on short notice without prior agreement between driver and passenger (Agatz et al., 2012). It doesn't require new network infrastructure and it is more convenient than public transport. Thus, projects that worked on mobile phones with Internet access and offered dynamic ridesharing applications (Dibbern, 2011) became feasible. They allowed drivers with spare seats to connect to people wanting to share a ride. At the same time, the safety of travelling with a stranger was increased through the use of reputation systems (e.g., PickupPal) or social networking tools (e.g., GoLoco and Zimride). It led to the new stage of ride-pooling service, as the high density and mobility of the population in cities made it possible for TNCs to combine travellers' trips in greater numbers, reducing their discomfort from additional waiting time and providing the desired discounts for the service provided (Alonso-Mora et al., 2017; Kaddoura & Schlenther, 2021; Zwick et al., 2021).

Thus, the potential benefits of the ride-pooling service have created a solid ground for the rapid development of the ride-pooling market: from the early test case of the Kutsuplus service "a convenient route for you, always" in 2012 (Rissanen, 2016) to the peak of development in 2019 when over 130 new services were launched across the globe (Foljanty, 2020). Ride-pooling services in many cities around the world became increasingly popular, as they offered convenient and cost-effective means of personal mobility (Fielbaum et al., 2021; Zwick & Axhausen, 2022).

*1.2.3 Pandemic disruption*

Yet, the COVID pandemic was a major disruption to the global transport industry in general (Borkowski et al., 2021), d the ride-pooling service in particular. Following government safety orders due to the COVID pandemic, travellers have reduced significantly their mobility. During the pandemic two major TNCs like Uber and Lyft experienced a 90% drop in rides in the USA, forcing operators to look for new business models to generate additional revenue (Curry, 2022; Iqbal, 2022). For ride-pooling service, it was even more disruptive, since it virtually ceased to exist. Given the significant risk of the virus spreading in ride-pooling networks (Kucharski, 2021), Uber (Iqbal, 2022) and Lyft (Macdonald, 2022) deactivated the ability to book pooled rides (Bursztynsky, 2020).

### 1.3 Prospects

Even so, despite temporal virus awareness, the general attractiveness of pooling, both for travellers and the platform providers remains intact. This is evidenced by the fact that the significant number of pilot ride-pooling projects continued to rise, even during a pandemic (Foljanty, 2021). After a pandemic pause, big TNCs also began to recover from the pandemic and declare their intentions to provide the ride-pooling service to the market. For instance, Lyft restarted service in Miami and Philadelphia in November 2021 (Hawkins, 2020). In May 2022, it announced plans to expand to San Francisco, San Jose, California, Denver, Las Vegas and Atlanta. Uber also announced the return of ride-pooling service in U.S. cities like New York and San Francisco in 2022 (Bursztynsky, 2022).

But can travellers really use this service today, as providers announce? Will they use it and recover the critical mass lost due to the pandemic? To date, those questions remain open. This, coupled with the uncertainty of travel behaviour after the pandemic, makes it increasingly ambiguous to assess the state of the ride-pooling market in the postpandemic period. To contribute to this, here we try to provide more precise answers to the following questions: what is the future of ride-pooling and how has the COVID pandemic affected it; whether it has recovered and reached the pre-pandemic levels; whether it will gain the critical mass needed to achieve significant market shares.

## 2. Ride-pooling market evolution: from the heyday in the pre-pandemic period to the restart in the postpandemic period

To answer the research questions we conduct three surveys: 1) we start with a literature review, giving a broad and deep overview of the ride-pooling market with prospects for its future growth; 2) then we report the unsuccessful survey that we addressed to ride-pooling professionals intending to receive their opinions on the future of ride-pooling; 3) to observe the postpandemic ride-pooling availability we investigate 5 ride-pooling apps in over 30 cities worldwide, to see whether the ride-pooling options were available in practice. Then, we summarise the reported changes in postpandemic travel behaviour and identify how the safety concerns of individuals and public decision-makers may drive the future of ride-pooling. Such a comprehensive approach allows us to conclude future ride-pooling prospects based on data available at hand.

### 2.1 Literature review

The evolution of the modern ride-pooling service is driven by the TNCs' search for a transportation service that will be a more attractive alternative for travellers. The introduction of big data analytics and artificial intelligence based on a mobility paradigm "mobility as a service" (MaaS) enabled the emergence of shared mobility online services (Matowicki et al., 2022; Schikofsky et al., 2020; Shaheen & Chan, 2016). TNCs have developed digital platforms as a means of connecting demanders and suppliers, which have enabled the emergence of ride-sourcing, ride-hailing and ride-pooling services not only in the mature transport markets of developed countries (Zwick & Axhausen, 2022) but also in emerging markets (Foljanty, 2020). With a growing number of service providers and an increasing number of travellers, ride-pooling could become efficient and sustainable, offering the travellers a service that was perceived as more attractive than the non-shared, private, alternative (Kucharski & Cats, 2020, 2022a). All of this pressured the taxi service, as the market expansion of such a corporation as Uber went global (Wong et al., 2020). However, TNCs lately seek to work with city governments and complement public transport through direct collaboration or as contract partners (Boone et. al., 2018; Zhu et. al., 2021). But without owning their fleet and without treating their driver/partners as employees, TNCs are potentially dangerous competitors to both public transport and the taxi industry. So, this forces the authorities to control and regulate the activities of TNCs. In many countries, the operations of app-based mobility services are not regulated by the government, so drivers can work without separate licences and registration. For instance, in 2021, Didi Global Inc.'s business declined sharply after regulators clamped down on its use of consumer data, resulting in a $4.7 billion loss (Savov & Liu, 2021). And regulators worldwide continue to adopt requirements regarding the use, transfer, security, storage, and other processing of personally identifiable information and other data relating to individuals (Ride Sharing Market, 2021).

Being an intermediate mode between individual rides and public transport, ride-pooling since its inception tried to keep the advantages of both private rides and mass transit. Private rides of ride-hailing services appeal to travellers as door-to-door services but have a much higher cost than the public transport option. Therefore, ride-pooling offers the traveller to "share" her trip with a companion(s), allowing for a slight increase in passenger hours (detour and delay due to pooling), which is compensated with a discounted fare.

So, what is the potential of ride pooling and why is it appealing for all the parties involved? First of all, ride-pooling service is a possible way of reducing the negative impact of private urban mobility (Kucharski & Cats, 2020; Santi et al., 2014; Zwick & Axhausen, 2022; Zwick et al., 2021). A potential reduction in the total number of trips due to their combination leads to a reduced vehicle usage time, thereby reducing vehicle kilometres travelled and required fleet sizes compared to single-passenger mobility options (Zwick & Axhausen, 2022). For instance, in the case of Amsterdam, average occupancy of 1.67 is yielding a 30% vehicle-hours reduction (Kucharski & Cats, 2020). Such compactness of pooling can thus contribute to traffic congestion and air pollution reduction. Notably, the overall passenger utility may still increase, due to the discounts offered, compensating for the longer travel times and less comfort (Alonso-Mora et al., 2017; Ilavarasan et al., 2019; Kucharski & Cats, 2020). But it needs a certain adoption rate. For example, a simulation study quantifying the benefits of autonomous on-demand ride-pooling for Munich (Germany) showed that pooling makes up for the additional empty vehicle miles travelled already starting from a 5% adoption rate (Engelhardt et al., 2019). Secondly, an obvious benefit of ride-pooling is the benefit to transport companies and their drivers, which by reducing vehicle kilometres travelled and required fleet can reduce their costs, thereby increasing their profits (Dandl et al., 2021; Kaddoura & Schlenther, 2021).

Undoubtedly, the great development of shared mobility online services has become possible thanks to algorithmic improvements which allowed a real-time solution to the highly challenging and computationally exploding ride-pooling problem (Agatz et al., 2012). Starting with a shareability network and a methodology to match travellers in pairs (Santi et al., 2014), up to the complete algorithm to efficiently match incoming requests with available vehicles (Alonso-Mora et al., 2017) which made it possible to propose real-time the pooled rides acceptable for travellers and optimal for ride-polling organisations while minimising service costs. In parallel, the emergence of two-sided mobility platforms has opened up new opportunities for studying the complex dynamics of platform-driven urban mobility. Researchers have developed open-source simulation frameworks (Kucharski & Cats, 2022b) to model two-sided mobility platforms and dynamic interaction between supply and demand (Seo & Asakura, 2022); travellers' mode and platform choices (Abouelela, et al., 2022); strategic decisions of drivers about participation, work shifts (Ashkrof, et al., 2020, Di et al., 2022); computational and organisational challenges of pooled rides (Fielbaum et al., 2021, 2022; Zwick et al., 2021); control of shared autonomous vehicles (Engelhardt et al., 2019; Seo and Asakura, 2022; Tian et al., 2022), etc. Thus, a series of improvements to the original solutions of recent developments in ride-pooling research has underpinned the ride-pooling services with an algorithmic base needed to start reaching the critical mass.

### 2.1.1 The covid disruption

However, the COVID pandemic has brought enormous impacts and changes to human mobility. Most countries have adopted various measures to limit the spread of the coronavirus, including lockdown, social distancing, reduction in population mobility, etc., which were primarily reflected in the transport sector, predominantly on using public transport (Agyeman et al., 2022; Barnes et al., 2020; Mathijs de Haas, et al., 2020). As a result, urban mobility has significantly declined around the world (MapBox, 2020; Rusul et al., 2022). For instance, in Sweden, there has been a 40%–60% decrease in public transport ridership across regions. The ridership reduction stemmed primarily from a lower number of active public transport travellers (Jenelius & Cebecauer, 2020). In Spain, the analysis revealed an overall mobility fall of 76%, and public transport users dropped by up to 93% (Aloi et al., 2020). Global quarantine and the transition of many businesses to a remote work option (for instance, in the Netherlands the share of workers that work completely from home increased from 6% to 39% (Mathijs de Haas et al., 2020) significantly reduced general mobility of people and the number of trips to the office (Savov & Liu, 2021). Such numbers and trends could be found in all the cities and countries facing consecutive waves, lockdowns and variants.

During the pandemic, similar to public transit, the taxi and ride-sourcing market also suffered much as a result of reduced travel demand ( Hu and Schweber, 2020; Yu et al., 2022).

Decreasing the demand for public transportation to avoid the risk of contracting Covid-19 could contribute to an increase in demand for the ride-pooling service, which, under specific control measures, can be a safe alternative in terms of the spread of viruses (Kucharski et al., 2021) and more efficient than a private car or taxi service. However, the ride-pooling market has become one of the most impacted transportation industries (Agyeman et al., 2022). Transportation network companies, such as Uber and Lyft, were also facing tremendous difficulties and were forced to deactivate their ride-pooling service (Iqbal, 2022; Macdonald, 2022).

Ride-hailing companies began to develop e-commerce, food delivery and last-mile delivery sectors (Curry, 2022) to survive. At the same time it is interesting to note that while major transportation network companies such as Uber and Lyft suspended their ridesharing services in March 2020, the number of new ride-pooling launches has continued to grow (more than 138 new projects for three quarters of 2021), even more than in 2019 in terms of the number of new services (Foljanty, 2021). Which gives hope that the ride-pooling market will not be completely closed and will start to function with renewed vigour in the postpandemic period.

In the context of the pandemic, while public transport inherently has a potential role in virus spreading (Gkiotsalitis & Cats, 2021), the introduction of an effective control measure allowing to halt the spreading before the outbreaks give the ride-pooling service another advantage - the potential of ride-pooling rides to serve as a safe and effective alternative given the personal and public health risks considerations. A study conducted in Amsterdam (Kucharski et al., 2021) showed that applying relatively straightforward control measures allows halting the virus spreading before the outbreaks without sacrificing the efficiency achieved by pooling. Starting from 20 initial spreaders the outbreak may reach up to 800 out of 2000 travellers, yet active control measures in the ride-pooling system made it possible to reduce it to 50 infections instead of 800. Such technological controllability gained incredible value during the COVID-19 pandemic when the world was faced with restrictions on transport and mobility at an unprecedented scale. Despite this, in practice, this opportunity was not exploited, and pooling was not available throughout the pandemic.

### 2.1.2 Postpandemic recovery stories

In turn, given the above advantages of the ride-pooling service and the positive dynamics of ride-pooling market development before the pandemic, some transportation network companies are restarting the ride-pooling service (Bursztynsky, 2022), and some transportation network companies continue to develop and release new projects (Foljanty, 2021). But the vast majority of new services are pilot projects. Turning an idea into a business is where the real challenge begins. Therefore, the significant number of ride-pooling providers cannot be indicative of its postpandemic recovery.

Considering that the share of pooled rides was found to be below 30% before the pandemic (Zwick & Axhausen, 2022) and was almost completely deactivated during the pandemic, the recovery period for the ride-pooling service after the pandemic is likely to be even longer compared to the rate of restoration of public transportation.

After announcing the restart of the ride-pooling service in November 2021 (Hawkins, 2020), Lyft faced problems both internal and external: an ongoing driver shortage, high gas prices and, as a result, low demand for the service. While high gas prices are a common problem for ride-pooling companies and cause financial hardship and a reduction in the number of drivers employed by TNCs (Campbell, 2022), there are also internal company problems. Providing the lowest fares for ride-pooling service before the pandemic did not make drivers willing to fulfil this kind of order (Astoria, 2022). The change in route and increase in travel time encouraged travellers to give low ratings to drivers. As drivers fled the platform, wait times increased, and the cost of rides soared. To counteract this, Lyft is developing a policy to make the service better and more reliable for passengers and drivers alike. By restarting the service, the company gave the drivers the right to decide on consent to perform shared rides. But the question of whether it became possible for the traveller to use the shared rides remains open.

Another large ride-pooling provider Uber has also faced similar problems in addition to the leaked "Uber files" scandal (Deslandes, 2022). Uber is still the market leader in the rideshare industry. However, following a passenger-centric course, alienates its drivers and makes it more difficult to refocus the business to keep drivers on the platform. The company tries to spend more on encouraging drivers and attracting demand while losing the stability of its work. To attract demand, the company introduced an upfront discount and up to 20% off the total fare if riders are matched with a co-rider along the way (Bursztynsky, 2022). At the same time, the company announced a limit level of increase in travel time for pooling passengers of 8 minutes. The company has limited rides to a total of two passengers. It expects a significant expansion of the ride pooling market in the postpandemic period.

Based on (Foljanty, 2021) in total, over 620 on-demand ride-pooling services have been deployed globally of which over 450 were running at the end of 2021. Apart from restarting the once existing ride-pooling services, new services continue to be opened from existing providers in the ride-pooling market. For instance, the new on-demand bus service FluxFux between Grevelsberg and Breckerfeld (Germany) was launched in July 2022 by the "door2door" provider (Foljanty, 2022a).

To complete the picture we note the significant paradigm shift in the recent financial markets. After many years of so-called "cheap money" that was flowing towards high-tech prospective ventures the current economical situation, triggered by barbarian Russian aggression towards the independent Ukraine nation, led to inflation and higher interest rates. This, in turn, led to an outflow of money from the market, in particular from high-risk tech start-ups and giants (like Uber) (Levi & Novet, 2022). We are afraid that this may significantly affect the financing of future start-ups, whose potentially successful ride-pooling business model may struggle to find financing.

### 2.2. Surveying ride-pooling professionals

To find out more about how ride-pooling can recover after the pandemic and develop further, we surveyed the target group of employees of companies who were or are currently interested in providing a ride-pooling service. Contacts of the target group were obtained primarily from LinkedIn, profiles of speakers of Conferences and Scopus papers. The number of respondents was 78. The questionnaire was designed with Google forms. It was distributed through emails and social media channels such as Facebook, LinkedIn, ORCID, and ResearchGate in June 2022. The questions (listed in Table 1) were about the prospects of introducing ride-pooling (Shulika & Kucharski, 2022).

Unfortunately, almost none of the respondents filled out the survey. This, as we believe, was due to two factors: a) the target group was very specific and small - ride-pooling professionals, presumably below 1000 competent respondents worldwide b) in the light of the recent Uber scandal the privacy and confidentiality issues of commercial providers prevented them from replying with details on future business strategies.

Thus, we just briefly synthesise the responses that we managed to collect, more like interesting voices rather than representative findings. Respondents believed in the promising prospects for the development of the pooling services. Insufficient demand (not enough riders to reach critical mass) and the need for a significant fleet size to be profitable was identified as the biggest obstacles to the introduction of ride-pooling.

**Table 1. The questions of the ride-pooling survey**

| Question | Type of a question | Answer options |
|---|---|---|
| Company Name (optional) | Paragraph | - |
| Country | Paragraph | - |
| In which regions does your company provide ride-pooling services? Please write down the names of regions, countries or cities. | Paragraph | - |
| Is your company providing now (i.e. postpandemic) the ride-pooling services? | Multiple choice | YES; NO |
| Did you stop providing ride-pooling during the COVID-19 pandemic? | Multiple choice | YES; NO; We did not provide ride-pooling pre-COVID |
| Is your company going to introduce ride-pooling services after the COVID pandemic? | Multiple choice | YES; NO; We have already provided it |
| Is your company's management satisfied with your ride-pooling business? | Linear scale | 1-Unsatisfied, not profitable<br>5 - Highly satisfied, profitable |
| What do you perceive as the biggest obstacles to introducing ride-pooling (select up to 4 options) | Checkboxes | ● Algorithms and operational planning);<br>● Management (real-time operations);<br>● Insufficient demand (not enough travellers to reach critical mass);<br>● Required fleet size (too many vehicles needed to reach profitability).<br>● COVID-related concerns (virus spreading);<br>● Marketing (travellers are not familiar with ride-pooling);<br>● Non-optimal business strategy (my company can earn more from ride-hailing);<br>● Users' reluctance (travellers will rather prefer individual rides);<br>● Drivers' reluctance (drivers working for my company are not interested in ride-pooling);<br>● Other |
| In your opinion: Will ride-pooling eventually gain the critical mass needed to achieve significant market shares? | Multiple choice | YES, in the short term (max. 10 years from now); YES, but in the longer term; NO |
| Do you think that ride-pooling can eventually become a profitable business? | Multiple choice | YES; only with public subsidies; NO |
| In your opinion: How successful will ride-pooling be in the future? | Linear scale | 1-Highly risky, unprofitable<br>5 - Stable and highly profitable |
| Can ride-pooling (in your opinion) become popular and frequently used among travellers? | Linear scale | 1-Rarely used, unpopular<br>5 - Frequently used, popular |
| Tell us more - what do you think about the perspectives for providing ride-pooling services | Paragraph | - |

*2.3. Empirical pooling availability study*

In this stage of analysis, we aimed to empirically verify the actual availability of ride-pooling services in the postpandemic period. Before the pandemic, shared rides were available in many cities and via several platforms. But while just about every other mode of transportation has returned after the pandemic, there are opinions that ride-pooling Uber (UberX) and Lyft (Lyft Shared) have been nowhere to be found (Desai, 2022). To verify, we examined the actual possibility of booking the service via a smartphone app by an ordinary traveller.

We carried out the service booking analysis with the following restrictions. First, according to the analysis of the ride-hailing market, the list of the largest transportation network companies (Curry, 2022) we selected the following biggest platforms: Uber, Lyft, DiDi, Ola, Grab, Bolt, Free Now, Cabify, Gojek and Gett. Among them, Uber (Iqbal, 2022), Lyft (Iqbal, 2022), DiDi (Savov & Liu, 2021), Ola (Vardhan & Tyagi, 2021) and Grab (Business Wire, 2022) were taken into account as they provided ride-pooling services before the pandemic. The data of the main companies offering successful ride-pooling services in the period before the pandemic is given in Table 2.

Secondly, we analysed the availability of ordering the ride-pooling service in 29 subjectively most important metropolises in the world (Figure 2). We report the following: name; the possibility of app installation on different operating systems; the name of the ride-pooling service and the territory where the ride-pooling service was provided.

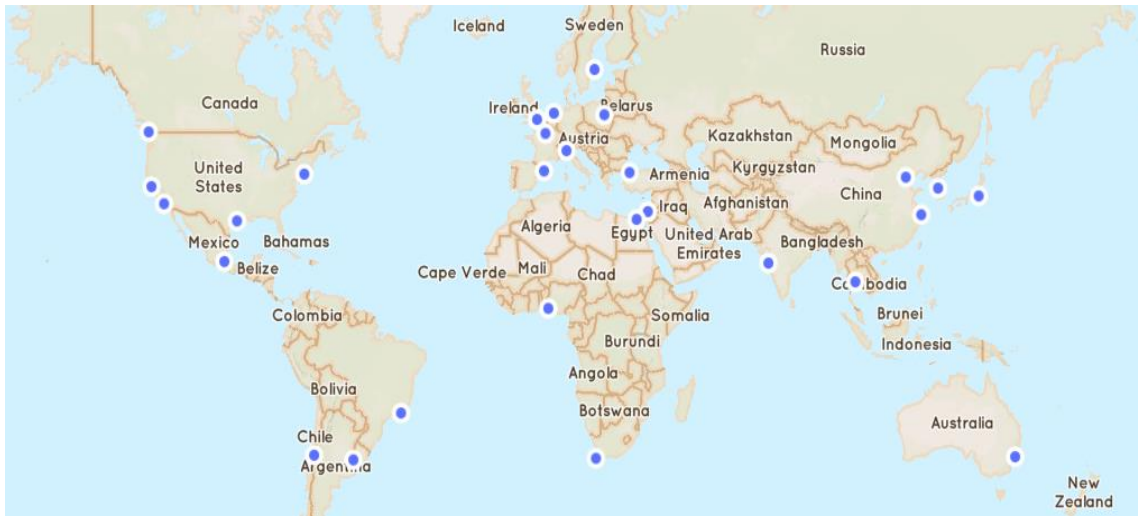

Figure 2. Mapping metropolises selected for the analysis

Based on that, we have checked the availability of the service (RH - ride-hailing and RP - ride-pooling) as of June 15, 2022, in the specified cities for the selected companies. We installed the apps of the selected ride-pooling companies on local phones, sometimes this required using VPN and registering via local services to verify identity. Then we queried several representative trip requests (e.g. from railway station to the centre, from airport to shopping centre, etc.) at various times of the day. We observed what kind of services were available, with a focus on postpandemic ride-pooling services.

**Table 2. The data of the main companies offering ride-pooling services before the pandemic**

| Company name | Name of app | The operating system of a smartphone | | Location | Service name |
|---|---|---|---|---|---|
| | | Android | iOS | | |
| Uber Technologies, Inc. (Curry, 2022, Iqbal, 2022) | Uber | + | + | USA, Europe, Middle East, Africa, more than 10,000 cities | UberPOOL |
| Lyft, Inc. (Iqbal, 2022) | Lyft | + | + | USA, Canada | Lyft Shared, Shared Saver |
| DiDi Global Inc. (Savov & Liu, 2021) | DiDi | + | + | China, Brazil, Mexico | DiDi Share, DiDi Hitch |
| Ola Cabs (Vardhan & Tyagi, 2021) | Ola | + | + | India, Australia, New Zealand and the United Kingdom | Share |
| Grab Holdings Inc. (BusinessWire, 2022) | Grab | + | + | Southeast Asia | GrabShare |

Unlike the announced return of the ride-pooling service to the global market, the analysis data showed that at the time of the analysis, only isolated cases of the provision of this service were recorded (Table 3). In the vast majority of cities where ride-pooling was provided before the pandemic, there is currently no opportunity to use it. Thus, the difference between the statements of transport companies about restarting ride-pooling in the postpandemic period and the real opportunities for travellers to book this service is more than palpable.

**Table 3. Data on the availability of ride-pooling service after the pandemic**

| City | Company name | | | | | |
|---|---|---|---|---|---|---|
| | Uber | | Lyft | DiDi | Ola | Grab |
| | RH | RP | RP | RP | RP | RP |
| Tokyo | + | - | | - | | - |
| Seoul | + | - | | | | |
| Sydney | + | - | | - | | |
| Cape Town | + | - | | | | |
| Lagos (Nigeria) | + | + | | | | |
| Cairo | + | + | | | | |
| Tel Aviv | | | | | | |
| Mumbai | + | - | - | | - | - |
| Bangkok | | | | | - | - |
| Singapore | | | | | | - |
| Istanbul | + | - | | | | |
| Warsaw | + | - | | | | |
| Amsterdam | + | - | | | | |
| London | + | - | | - | - | - |
| Paris | + | - | | | | |
| Barcelona | - | - | | | | |
| Milan | + | - | | | | |
| Stockholm | + | - | | | | |
| New York | + | - | - | - | - | - |
| Los Angeles | + | - | - | - | - | - |
| San Francisco | + | - | - | - | - | - |
| Vancouver | + | - | - | - | - | - |
| Houston | + | - | - | - | - | - |
| Mexico City | + | - | - | - | | |
| Rio De Janeiro | + | - | | | | |
| Buenos Aires | + | - | | - | | |
| Santiago | + | - | | | | |
| Shanghai | + | - | | - | | |
| Beijing | | | | - | | |

## 2.4 Traveller behaviour

Since users' sociodemographic and travel characteristics of travellers are the main drivers to shift to pooled rides (Abouelela et al, 2022), the objective of the following section of our analysis is to identify trends in travellers' behaviour and attitudes toward ride-pooling services. In this section, we report the state-of-the-art in traveller behaviour in the context of ride-pooling. We report what behaviour and attitudes led to the rise of ride-pooling, how COVID affected travel behaviour and if COVID was disruptive enough to induce long-standing changes in travel behaviours, which would prohibit recovery of ride-pooling services.

The daily behaviour of travellers on trips is largely dependent on habits and daily routines (Schönfelder & Axhausen, 2010). Therefore, changes in behaviour do not occur often. A significant event is needed that will entail a change in the way people live. When there are significant changes in the status of public transport systems, understanding user perceptions are of key importance (Allen et al, 2019).

### 2.4.1 Pre-pandemic

In 2019 over 130 new services were launched across the globe (Foljanty, 2020) and the service became increasingly popular in many cities around the world. But has this service reached the critical mass needed to achieve significant market shares? Analysis of studies did not give an unambiguous answer, but we note that the use of ride-pooling services has been rather low (Gehrke et al., 2021). In cities where pooled alternatives were available, only 20% of ride-hailing trips were pooled (Alonso-González et al., 2021).

Therefore, many researchers (such as Alonso-González et al., 2021; Gehrke et al., 2021; Geržinič et al., 2022; König & Grippenkoven, 2020; Wang & Noland, 2021 etc.) have become interested in getting a better understanding of travellers' preferences to make ride-pooling more attractive since the traveller's willingness to share his/her rides with other passengers for a cheaper fare determines his/her choice of individual or pooled service (Alonso-González et al., 2021). To do this, some studies (Liu et al., 2019; Yan et al., 2019 etc.) used stated preference experiments (typically applying a multinomial logit model or a mixed logit model formulation) or through different latent class clustering methods (Alemi et al., 2019; Geržinič et al., 2022 etc.). Geržinič et al. (2022) provide an overview of studies using stated preference data collection. Thus, the predominant number of studies on the main factors that affect the willingness-to-share of travellers highlight the cost of travel (Alonso-González et al., 2021; de Souza Silva et al., 2018; Gehrke et al., 2021; Geržinič et al., 2022; Morales Sarriera et al., 2017; Spurlock et al., 2019; Wang & Noland, 2021), travel time (Alonso-González et al., 2021; Geržinič et al., 2022; Wang & Noland, 2021), waiting time (Alonso-González et al., 2021; Geržinič et al., 2022), safety and information about future passengers (race, gender) in the vehicle (de Souza Silva et al., 2018; Morales Sarriera et al., 2017). At the same time, the willingness to share trips differed across different demographics, socioeconomic status travellers, and built environment characteristics (Alonso-González et al., 2021, Gehrke et al., 2021; Spurlock et al., 2019; Wang & Noland, 2021). The results of the analysis showed that in the pre-pandemic period both researchers and TNCs identified critical factors of human behaviour that allowed the ride-pooling services to take off. Also, the introduction of new technologies made it possible for TNCs to offer lower fares with door-to-door ride-pooling characteristics. This favourably influenced the willingness of travellers to use the service of ride-pooling, as a result of which ride-pooling became an increasingly popular service.

### 2.4.2 Pandemic

Meanwhile, the pandemic has brought significant lifestyle changes to people around the world. As a result of government measures introduced during the COVID-19 pandemic for the sector transport safety, people reduced their activities outdoors (Jenelius & Cebecauer, 2020). In the Netherlands this reduction was approximately 80% Mathijs de Haas et al. (2020); more than 40% of workers increased the number of hours working from home (27% of home-workers expect to work from home more often in the future). In the megacity of Istanbul (Turkey), the demand for public transport has fallen by more than 85% during the pandemic (Aydin et. al., 2022). Notably, the pandemic led to some changes in travellers' travel mode choices. For instance, Bucsky, (2020) explored a decrease in demand for public transport by about 80% and an increase in car use from 43% to 65% in Budapest, Hungary.

Consequently, the daily behaviour of travellers began to change. According to Abdullaha et al., (2020) the primary purpose of travelling has shifted away from work (a decrease from 58% of the respondents before COVID-19 to only 30% during COVID-19). Shopping became the primary purpose of travelling (increase from 4% to about 44% of the respondents during COVID-19). 36% of respondents stated that they used public transport for their main purposes before COVID-19 and only 13% during COVID-19. In contrast, private car use increased from 32% before COVID-19 to 39% during COVID-19.

Unsurprisingly, during the pandemic, respondents paid more attention to factors related to the risk of infection: safety, cleanliness, social distancing, wearing masks, online prepaid fare system, door-to-door delivery, etc. The measures introduced during the COVID-19 pandemic for the transport sector on safety and reliability reinforced the perceptions of service quality by travellers, particularly concerning comfort and crowding (Esmailpour et al., 2022). At the same time, much less attention was paid to cost and travel time savings. And this already indicates a significant change in traveller behaviour. Due to the risk of being infected, people were unwilling to take public transportation in preference for private cars and cycling as a safe way of travel (GehrAbdullahke, 2021; Ferreira et al., 2022; Vinod & Sharma, 2021).

All of these results show that the pandemic led to a change in peoples' behavioural habits, in particular, it dramatically changed the perceptions toward ride-pooling. People were scared of COVID - stayed home and avoided crowds (including ride-pooling, commonly perceived as risky). People were forced to follow directives to stay at home, many began to form new habits, with new behaviour. These may affect the behaviour long after the direct virus threat is no longer real, which will have a significant impact on traveller behaviour concerning the ride-pooling service as well.

*2.4.3 Postpandemic*

The analysis of traveller behaviour now, in the postpandemic period, is of particular interest. Opening up this issue is not so easy, as only recently the transport sector has experienced a long period of pandemic.

It is still unpredictable if and when the transport sector will recover to the pre-pandemic levels, however a lower monetary cost compared to a personal car and the end of quarantine restrictions by many states contributed to a gradual restoration of public transport in the postpandemic period (Ferreira et al., 2022). The provided safety measures in public transport have led to eliminating passenger anxiety and restoring their confidence in public transport after a major safety crisis (Dong et al., 2021; Ferreira et al., 2022; Jenelius & Cebecauer, 2022). We can observe that public transport is full again. For instance, the number of active travellers per day in Stockholm public transport has now recovered to more than 80% of pre-pandemic levels (Jenelius & Cebecauer, 2022). By 2023 (or by 2025 due to the crisis caused by Russia's attack on Ukraine), the Netherlands government predicts a full recovery of public transport use at 2019 levels (Francke & Bakker, 2022).

In the On-Demand Transit services market, according to Lukas Foljanty (2022b), in 2022, a good pace of entry into the market and development of new On-Demand

Transit services (it was about 40 launched in Q2 2022) is maintained, with a continued decline in B2C Ride-pooling services since 2019.

Therefore, given that people seem not to be scared by the threat of infection anymore and TNCs are interested in the recovery of the ride-pooling market, the postpandemic ride-pooling market looks like a lively space for new developments and promotion of existing projects.

## 3. Postpandemic ride-pooling market prospects

Gaining good momentum in the pre-pandemic period, having experienced a virtual cessation of activity during the pandemic, the postpandemic ride-pooling market needs to be carefully analysed and prospects for further development need to be determined. Our analysis aimed to show the main technological and social trends in the postpandemic ride-pooling market.

The first aspect is the composition of the participants in the ride-pooling market after the pandemic. Integral participants in the postpandemic ride-pooling market remain the travellers and transportation network companies/drivers. But the qualitative and quantitative characteristics of the participants and their relationship with each other are in constant change. Around the world, the governments were introducing strict regulations to improve traveller safety, continuing to increase the influence and control over the activities of transportation network companies, often hindering the development of the market and limiting the activities of TNCs. However, companies are implementing security measures for building trust in services in two major senses: there is no increased risk of infection and there is no risk of crime by drivers.

The key industry players have not changed in the postpandemic period. These remain TNCs giants such as Uber, Lyft, Didi, and Ola. Although the pandemic had a significant negative impact on the entire transport sector, including large TNCs, they survived in the market by focusing on providing a unique strategy. As these TNCs have community-based platforms that can revamp their business structure to have a competitive advantage, the companies have managed to remain a cost leader or provide differentiation. On the other hand, focusing major players on the development of the company in key areas creates opportunities for faster and more flexible start-ups, offering new ride-pooling services. Most likely it would be difficult for small players to capture the operational effectiveness and the established companies will be able to further dictate the ride-pooling market. However, all this shows the continued development of the ride-pooling market in the postpandemic period.

When analysing the state of the market, we paid special attention to another important issue both before and after the pandemic. This is the implementation of a competent policy of TNCs concerning drivers. As the experience of such a large TNC as Uber has shown, a customer-oriented strategy can lead to a mass refusal of drivers from work (Campbell, 2022), therefore all TNCs must take all necessary measures to ensure that the ride pooling service is attractive not only for travellers but also for drivers.

The next aspect of our analysis is the availability of ride-pooling service for travellers during the postpandemic period. We analysed the availability of ordering the ride-pooling service in 29 most important representative metropolises in the world. We took into account service providers such as Uber, Lyft, DiDi, Ola and Grab. Instead of the announced return of the ride-pooling service to the global market, we marked that only isolated cases of the provision of this service were available. In the vast majority of cities where ride-pooling was provided before the pandemic, there is currently no opportunity to use it. Thus, there is a significant difference between the statements of TNCs about restarting ride-pooling in the postpandemic period and the real opportunities for travellers to book this service.

Another important aspect is the change in traveller behaviour. We reported significant changes in traveller behaviour in the postpandemic ride-pooling market. While before the pandemic fare was one of the central drivers catalysing the shift towards pooled services, during the pandemic travellers were more concerned about the safety and prevention measures taken. Now travellers seem not to be so concerned by the threat of infection anymore. Yet, there is another factor that makes it much more difficult for the ride-pooling market to quickly recover to its previous level. People were forced to follow directives to stay at home, many began to form experiences with new behaviour. Having gained remote work experience, people might for instance prefer to work from home. While all this, online shopping, and more accurate information about crowds and the attractiveness of specific places don't necessarily mean that people travel less, they do travel to a wider range of different destinations. Which can become a new market niche for pooled services, though challenging due to lower demand levels and spatially scattered destinations.

Also, to find out how ride-pooling can recover after the pandemic and develop further, we surveyed the target group of employees of companies who were or are currently interested in providing a ride-pooling service. The questions were about the biggest obstacles to introducing ride-pooling and its prospects. Unfortunately, due to the small number of survey participants who volunteered to answer our questions, the results of the survey cannot be considered representative.

## 4. Conclusion

The COVID pandemic had a big impact on the ride-pooling market, which is unlikely to recover to pre-COVID levels anytime soon. Though, the key upper hands of ride-pooling are a solid base for future growth. The ride-pooling affordable and seamless door-to-door services provide benefits that neither public transport nor ride-hailing can provide. And this trend is likely to get even stronger once ride-hailing services become autonomous and can operate at even lower prices.

An overview of the current state of the ride-pooling market shows the opportunities for the sharing economy to develop in the postpandemic situation. Our analysis suggests that TNCs are highly likely to continue to develop a ride-pooling service; new ride-pooling projects will continue to appear on the market in an attempt to express their uniqueness and travellers are highly likely to continue using the pooled rides with precautionary additions.

To conclude, the ride-pooling market still did not recover after the pandemic disruption. It was halted just when it started reaching critical mass and becoming increasingly popular. While most of the transport industry is at a fast recovery trend, with crowded public transport systems and congested motorways, ride-pooling is still not available in practice. Nonetheless, the core elements of the ride-pooling business model were not affected by the pandemic. It remains a promising option for all the parties involved, with a great potential to become attractive for travellers, drivers, TNC platforms and policymakers. The travel behaviour changes due to the pandemic seem not to be long-lasting, our virus awareness is no anymore the key concern and our willingness to share and reduce fares seem to be high again.

Yet, whether ride-pooling will get another chance to grow remains open. The number of launches of ride-pooling start-ups is unprecedented, yet the financial perspectives are unclear. Ride-pooling needs to be subsidised before reaching critical mass and becoming a sustainable business model. Growing interest rates makes it harder to finance, making investors impatient and looking for quick returns. Soon we will be able to verify whether some of those projects will manage to kick off and become the global ride-pooling actor. As we showed the pandemic obstacles are gone and key advantages of pooling remain in place.

**Disclosure statement**

No potential conflict of interest was reported by the author(s).

**Funding**

This research was funded by the National Science Centre in Poland program OPUS 19 (Grant Number 2020/37/B/HS4/01847).